\documentstyle[preprint,prl,aps,epsf]{revtex}

\begin{document}
\draft
\title{SUBDOMINANT INTERACTIONS AND $H_{c2}$ IN UBe$_{13}$.}

\author{I. A. Fomin$^1$, and J.P.~Brison$^2$}

\address{$^1$ P.L. Kapitza Institute for Physical Problems, Kosygin 
St.  2, Moscow 117334, Russian Federation, and
\\ D\'epartement de 
Recherche Fondamentale sur la Mati\`ere Condens\'ee, SPSMS, CEA-CENG, 
38054 Grenoble-Cedex 9, France\\
$^2$Centre de Recherches sur les Tr\`es Basses
Temp\'eratures, CNRS, BP 166, 38042 Grenoble-Cedex 9, France\\
}
\maketitle

\begin{abstract}
We discuss a model based on a field-induced mixture of two odd-parity 
irreducible representations to explain the unusual features of 
$H_{c2}(T)$ in the heavy fermion compound UBe$_{13}$.  We compare its 
predictions with recent pressure measurements as well as with the most 
prominent theoretical models which have been proposed up to now.
\end{abstract}
\bigskip
\pacs{PACS numbers:}

\section{Introduction}

Heavy-fermion compounds are well-known candidates for a search and 
investigation of unconventional superconductivity.  
In these three dimensional systems, the symmetry analysis of the 
possible superconducting state, depending on the crystalline lattice 
of the compound \cite{Gorkov87}, sets the frame for the identification of 
the unconventional superconducting phase.  This analysis gives a 
complete list of possible superconducting phases together with their 
properties determined by symmetry, including the type and order of 
nodes in the superconducting gap.  In spite of a favorable theoretical 
situation and numerous experimental work, there doesn't exist yet a 
firm and widely accepted identification of these superconducting 
phases (comparable to the results on superfluid $^{3}$He) in any heavy 
fermion superconductor.
  
In this paper, we examine the upper critical field $H_{c2}$ of the 
heavy fermion superconductor UBe$_{13}$ as a probe of the symmetry of 
its superconducting phase.  The upper critical field is usually not 
very sensitive to gap nodes (see for example the quantitative studies 
for various scenarios in UPd$_{2}$Al$_{3}$ \cite{Hessert97}).  But it 
is sensitive to the spin state of the Cooper pairs or more generally 
to the parity of the order parameter, because in heavy fermion 
systems, the orbital limitation is so large that $H_{c2}$ may be 
governed by the paramagnetic limitation.

UBe$_{13}$ is a cubic compound with a $T_{c}$ of order 1K, and it is 
a non-magnetic superconductor.  Since the first measurements 
of its upper critical field \cite{Maple85}, it is known that it 
presents two very unusual and intriguing features.  First, $H_{c2}$(T) 
has a strong negative curvature close to $T_{c}$ which changes sign at 
intermediate fields.  Then, taking account of a realistic value of the 
conduction electrons gyromagnetic $g-factor$, the paramagnetic limit 
at T=0 is exceeded several times, while the strong negative curvature 
close to $T_{c}$ shows that there exist a pronounced effect of the 
paramagnetic limitation.

Numerous explanations have been proposed since the first precise 
measurements of $H_{c2}(T)$ in this compound \cite{Maple85}.  Some 
have relied on additional hypothetic magnetic phase transitions 
\cite{Schmiedeshoff92}, or on the field dependence of the normal state 
properties \cite{Tachiki85,Rauchschwalbe85,DeLong87,Brison89}.  But 
none of these phenomenological interpretations have found a firm basis 
in other measurements or theoretical developments.  Another 
hypothesis, much closer to the point of view adopted here, relies on 
two different superconducting order parameters with a weak Pauli 
limitation \cite{Rauchschwalbe87}.  It has not been carried out 
quantitatively, but experimental support has been seeked through the 
detection of a possible second phase transition already in zero field 
below the main superconducting transition 
\cite{Rauchschwalbe87,Kromer98}.  To our opinion, such a second phase 
transition is not supported by the data, which only show a weak and 
smeared maximum in the specific heat or minimum in the thermal 
expansion.  The model that we propose here involves a mixture of two 
different irreducible representations, but does not rely on nor 
predict such a second phase transition.

At present, the only competing quantitative explanation relies on a 
simple strong-coupling model \cite{Thomas96}.  It has been also
successful in describing the evolution of both features up to 
pressures of 20 kbars \cite{Glemot99}: the complete temperature and 
pressure dependence of $H_{c2}$ comes out from a straightforward 
strong-coupling calculation with a single even parity state.  The 
conflict with the paramagnetic limit at T=0 is resolved by its 
enhancement due both to direct strong-coupling effects (increase of 
the ratio $\Delta / T_{c}$) and to the (parameter free) inclusion of the 
formation of a spatially modulated superconducting state (FFLO), 
induced by the dominance of the paramagnetic limitation.  
Very good agreement with the data of Ref. \cite{Glemot99} is provided 
in the whole pressure range, for $g$ being close to its free 
electron value, by fitting the strong coupling 
constant $\lambda$.  The pressure dependence of $\lambda$ agrees with 
that of the effective masses (as indicated by the Sommerfeld 
coefficient or the slope of $H_{c2}$ at $T_{c}$), but it also turns 
out to be exceptionaly large: $\lambda \approx 15$ at p=0. For all 
other known superconductors where the strong coupling regime due to 
electron-phonon interaction is well characterized, $\lambda$ does not 
exceed 5.

The model in this paper is based on a field-induced mixture of two 
odd-parity irreducible representations of the symmetry group ($O_{h}$) 
of the normal phase of UBe$_{13}$.  This possibility was already 
evoked in the literature \cite{Klemm85,Luk87}, but it has not been 
worked out to the extent that it could justify the particular choice 
of the order parameters and could be compared directly with the 
experimental data.  Here, the choice of the two representations 
is unique, thus providing a firm identification scheme for the 
superconducting state of UBe$_{13}$.  Within this scheme, the dominant 
component of the order parameter is analogous to the B-phase of the 
superfluid $^{3}$He.  We make a quantitative comparison of the 
theoretical predictions with the data of Ref.\cite{Glemot99} and propose 
an experimental check of the suggested scheme.

As we shall see, our choice of order parameters does not allow for a 
consistent interpretation of all data in UBe$_{13}$: in particular, 
it does not predict any nodes of the gap, whereas both old \cite{Ott87} 
and new \cite{Walti99} thermodynamic experiments hint at the 
presence of point nodes. We consider this approach as a first step 
which may help to clarify at least one aspect of the properties of UBe$_{13}$, 
bearing in mind that the complexity of the physics of heavy fermion 
systems rarely allows for fully satisfying explanations\ldots

\section{Choice of the representations.}

Rotational symmetry of the crystalline lattice of UBe$_{13}$ is 
described by the $O_{h}$ group.  Following the list of Ref.  
\cite{Volovik85} for strong spin orbit coupling, we have to consider 
10 irreducible representations: $A_{1 g,u}$; $A_{2 g,u}$; $E_{g,u}$; 
$F_{1 g,u}$; $F_{2 g,u}$.  The subscripts g and u denote 
correspondingly even or odd symmetry of the particular representation 
with respect to the inversion.  The even representations are the 
analogs of the spin singlet pairing and the odd - of the spin triplet.  
The order parameter for an even representation is a scalar and for the 
odd - a vector function of the direction in the momentum space 
$\vec{d}(\vec{k})$.  The capital letters denote A - one-dimensional, E 
- two-dimensional and F - three-dimensional representations.  To 
determine which of the representations is realized as the order 
parameter in a particular compound, one has to know the projections of 
the scattering amplitude $V_{\alpha \beta \lambda 
\mu}(\vec{k},\vec{k'})$ of quasiparticles in the normal phase on the 
basis functions of these representations.  The scattering amplitude is 
then represented in the form:
\begin{eqnarray}
	V_{\alpha \beta \lambda \mu}(\vec{k},\vec{k'})={1 \over 2}
	\sum_{\Gamma_{g}} g_{\beta \alpha}g^{+}_{\lambda \mu}V_{\Gamma_{g}}
	\sum_{i=1}^{d_{\Gamma}}\Psi_{i}^{\Gamma}(\vec{k})\Psi_{i}^{* 
	\Gamma}(\vec{k'}) \nonumber \\ 
	+ {1 \over 2}
	\sum_{\Gamma_{u}} V_{\Gamma_{u}}
	\sum_{j=1}^{d_{\Gamma}} (\vec{\Psi}_{j}^{\Gamma}(\vec{k}).\vec{g}_{\beta \alpha})
	 (\vec{\Psi}_{j}^{* \Gamma}(\vec{k}).\vec{g}^{+}_{\lambda \mu})
 \label{ScatAmp}
\end{eqnarray}                                   
The spin matrices are defined as $\vec{g}_{\alpha 
\beta}=(i\vec{\sigma}\sigma^{y})_{\alpha \beta}$, $g_{\alpha 
\beta}=i\sigma^{y}_{\alpha \beta}$ where $\sigma^{x}_{\alpha 
\beta}$,$\sigma^{y}_{\alpha \beta}$,$\sigma^{z}_{\alpha \beta}$, are 
Pauli matrices.  The sums are taken over all even and odd 
representations respectively.  The greatest of the positive amplitudes 
$V_{\Gamma_{0}}$ (the dominant one) determines the temperature of the real 
superconducting transition $T_{c0}$ as prescribed by the BCS-theory
\begin{equation}
T_{c0}=\frac{2\gamma}{\pi}\epsilon_{\Gamma}
\exp\left(-\frac{1}{N(0)V_{\Gamma_{0}}}\right)\label{Tc0}
\end{equation}
where $\ln\gamma$ is Euler's constant, $\epsilon_{\Gamma}$ a cut-off 
parameter and N(0) the density of states at the Fermi level.  All 
other amplitudes are subdominant.  The transformation properties of 
the order parameter not too far from $T_{c0}$ are described by the 
representation $\Gamma_{0}$ .  A magnetic field (H) changes the 
symmetry of the system and the classification of the representations.  
If only one, dominating term $V_{\Gamma_{0}}$ is kept in the sum 
(\ref{ScatAmp}) and if $\Gamma_{0}$ is a multidimensional 
representation, the magnetic field splits it.  Several branches 
$H_{c2}$(T) start in that case in the plane (T,H) from the point 
($T_{c0}$,0).  According to Ref.  \cite{Zhitomirskii92}, if the 
direction of the field H coincides with the symmetry axis of the 
lattice, the different branches are classified by two quantum numbers.  
The first is N=m+n, where m is the projection of angular momentum on 
the direction of field and n is the number of the Landau level, which 
describes the spatial dependence of the order parameter for the 
nucleating superconducting region.  The second quantum number is the 
parity $\sigma = \pm 1$ with respect to reflection in the plane 
perpendicular to the direction of H.  The branch with the highest H 
for a given T determines the true $H_{c2}$(T) \footnote{Strictly 
speaking even for conventional s-wave pairing, there are many branches 
$H_{c2}$(T) starting from $T_{c}$ and corresponding to different n, 
but one knows that the highest branch corresponds to n=0.}

The dependence $H_{c2}$(T) for p-wave pairing has been analyzed in 
detail by Scharnberg and Klemm \cite{Scharnberg80}.  No change of sign 
of the curvature was found, which is natural in a problem with only 
one parameter $V_{\Gamma_{0}}$.  But if all terms are retained in the 
sum (\ref{ScatAmp}), the magnetic field has an additional effect: it 
can mix basis functions having the same quantum number N, which 
originally belonged to different representations.  Substantial change 
of the original basis functions can take place for fields of order 
$\mu_{B}H \approx T_{c}$.  A change of the basis functions will change 
the projections $V_{\alpha \beta \lambda \mu}(\vec{k},\vec{k'})$ on 
these functions, including the projection which determines the 
superconducting transition.  A substantial admixture of the subdominant 
coefficients $V_{\Gamma}$ may therefore also take place on
energy scales $\approx T_{c}$, imitating an extreme strong coupling effect.  
As regard the field induced change of the interaction
potential itself, it is of the order of  $\mu_{B}H / \epsilon_{F}$ 
and can be neglected.

The minimum model which can include these effects must contain two 
representations well mixed by the magnetic field.  An advantage of the 
minimum model is that it remains tractable and contains only one 
additional parameter.  Whether or not the two-representations model is 
sufficiently accurate for a description of the magnetic properties of 
a particular superconductor depends on the values of coefficients in 
the sum (\ref{ScatAmp}) for this material, and on the temperature 
interval in which one expects to reproduce these properties.  In what 
follows, we apply the two-representation model to reproduce the 
unusual temperature dependence of $H_{c2}$ in UBe$_{13}$.  We use the 
above mentioned characteristic features of this dependence as a 
guidance for the choice of two dominating representations.

The critical magnetic field can exceed the paramagnetic limit if the 
order parameter in the high field region is predominantly of the 
odd-parity type (i.e.  a vector-function $\vec{d}(\vec{k})$), and if 
the projection of $\vec{d}(\vec{k})$ on the magnetic field is small or 
absent.  To provide a good mixing of the two participating 
representations by the magnetic field we assume that both 
representations are of the odd-parity type.  The pronounced 
paramagnetic effect in the low field region indicates that in that 
region the order parameter is dominated by a one-dimensional 
representation.  In that case it is not possible to eliminate 
completly one projection of $\vec{d}(\vec{k})$.  Such an elimination 
becomes possible in strong fields when the admixture of the second 
representation is appreciable.  An inspection of the representation 
table for the $O_{h}$ group with the corresponding basis functions 
\cite{Mineev99} suggests the following scheme: the superconducting 
phase which appears at H=0 and T=$T_{c0}$ belongs to the $A_{1u}$ 
representation with the basis function
\begin{equation}
\vec{\Psi}_{0}(\vec{k}) = \hat{x}k_{x}+\hat{y}k_{y}+\hat{z}k_{z}
	\label{A1u}
\end{equation}
The unit vectors $\hat{x}$, $\hat{y}$, $\hat{z}$ are directed along 
three mutually perpendicular four-fold axes of the cube.  The order 
parameter of the form (\ref{A1u}) corresponds to a state of spin S=1, 
orbital momentum L=1 and total angular momentum J=0.  This is the most 
symmetric odd-parity state, which is analogous to the B-phase of 
superfluid $^{3}$He.  The superconducting gap for that state has the 
full symmetry of the $O_{h}$ group and it does not have nodes required 
by symmetry.  We assume in what follows that the magnetic field is 
oriented along the z-direction.  Only one branch of $H_{c2}$(T) starts 
from $T_{c0}$ (ignoring higher Landau levels) and this branch is 
characterized by the quantum numbers $N=0$ and $\sigma=+1$.  The 
second representation is $E_{u}$.  For the present discussion it is 
convenient to choose the basis functions of $E_{u}$ in the following 
form:
\begin{eqnarray}
\vec{\Psi}_{1}(\vec{k}) &=& 
\frac{1+\epsilon}{\sqrt{2}}(\hat{x}k_{x}+\hat{y}k_{y}-2\hat{z}k_{z}),\label{Eu1}\\
\vec{\Psi}_{2}(\vec{k}) &=& 
\frac{1-\epsilon^2}{\sqrt{2}}(\hat{y}k_{y}-\hat{x}k_{x}),
\label{Eu2}
\end{eqnarray}
 where $\epsilon = e^{2\pi i/3}$ is a cubic root of 1.
${\Psi}_{1}(\vec{k})$ transforms as a
function with m=0, and  ${\Psi}_{2}(\vec{k})$ with $m=\pm 2$. One can see
immediately that the z-component of the combination
\begin{equation}
\vec{\Psi}_{\infty}(\vec{k}) = \sqrt{2}(1+\epsilon)\vec{\Psi}_{0} + 
\vec{\Psi}_{1},
	\label{PsiInfini}
\end{equation}
is zero and the paramagnetic limitation is absent for such an order 
parameter.  The elimination of the z-component would not be possible for 
the $A_{2u}$ representation, since as a function of $\vec{k}$ it is orthogonal 
to z-components of the basis functions of all other odd 
representations. So, we assume that the pairing potential contains 
contributions of only two representations $A_{1u}$ and $E_{u}$.
\begin{eqnarray}
	V_{\alpha \beta \lambda \mu}(\vec{k},\vec{k'})=
	{1 \over 2}V_{0}
	(\vec{\Psi}_{0}(\vec{k}).\vec{g}_{\beta \alpha})
	(\vec{\Psi}_{0}^{*}(\vec{k}).\vec{g}^{+}_{\lambda \mu}) \nonumber \\ 
	+ {1 \over 2}V_{1}
	\sum_{s=1,2}(\vec{\Psi}_{s}(\vec{k}).\vec{g}_{\beta \alpha})
	(\vec{\Psi}_{s}^{*}(\vec{k}).\vec{g}^{+}_{\lambda \mu})
 \label{ScatAmp2}
\end{eqnarray}
with two independent coupling constants $V_{0}$ and $V_{1}$ or,
formally, two BCS transition temperatures $T_{c0}$ and $T_{c1}$.
Then  the order parameter in the vicinity of the transition line
$H_{c2}$(T) is a linear combination of the basis functions
$\vec{\Psi}_{0}, \vec{\Psi}_{1}, \vec{\Psi}_{2}$:
\begin{equation}
\vec{d}(\vec{k},\vec{R}) = \Delta_{0}(\vec{R})\vec{\Psi}_{0}(\vec{k})
                           + \sum_{s=1,2}\Delta_{s}(\vec{R})\vec{\Psi}_{s}(\vec{k})
                          \label{d123}
\end{equation}
where $\Delta_{j}(\vec{R})$ j=0,1,2 are functions of the coordinates.

\section{Critical field $H_{c2}$(T)}
The dependence of $H_{c2}$(T) for our model can be found as in Ref.  
\cite{Scharnberg80}.  As a starting point we use Eq.(3) of 
Ref. \cite{Klemm85}.  With the interaction potential given by 
Eq. (\ref{ScatAmp2}) and the order parameter of Eq. (\ref{d123}), the 
equations for the functions $\Delta_{j}(\vec{R})$ are:
\begin{eqnarray}
 \frac{1}{N(0)V_{j}}\Delta_{j}(\vec{R})=2\pi 
 T\sum_{\omega_{n}}\sum_{j'=0}^{2}
 \int 
 \frac{d\Omega'}{4\pi}\vec{\Psi}_{j}^{*}(\vec{k'}).\int_{0}^{\infty}ds\nonumber \\
e^{-sL_{op}}.\{1-[1-\cos(2sgH)]\hat{n}\hat{n}^{tr}\}
\Delta_{j'}(\vec{R})\vec{\Psi}_{j'}(\vec{k'}),
\label{EqHc2}
\end{eqnarray}
where
\begin{equation}
L_{op}=2 |{\omega_{n}}| - 
i \mathrm{sgn}(\omega_{n})\vec{v}_{F}(k_{F}).(i\vec{\nabla}+2e\vec{A}(\vec{R}))
    \label{Lop}
\end{equation}
$j$~= 0,1,2; $\vec{A}(\vec{R})$ is the vector potential for the magnetic 
field H , $g$ is an effective gyromagnetic ratio. It need not be equal to its 
free electron value $g=2$: besides the usual renormalization due to 
spin orbit coupling, it includes here also possible Fermi-liquid 
corrections.In what follows, $g$ will be used as a fitting 
parameter.  The unit vector $\hat{n}$ is 
parallel to the H direction (thereafter also called z-direction).  
This is a linear system of differential equations of infinite order.  
Its solutions are given by the eigenfunctions $f_{n}(\vec{R})$ for the 
Landau levels of an electron under magnetic field:
\begin{equation}
\Delta_{j}(\vec{R})=\sum_{n}\eta_{jn}f_{n}(\vec{R})
\label{fLandau}
\end{equation}

Now we use the fact that the combination (\ref{d123}) corresponds to 
the quantum number N=0.  For the functions $\vec{\Psi}_{0}$ and 
$\vec{\Psi}_{1}$ m=0, while for $\vec{\Psi}_{2}$ $m=\pm 2$.  
This selects n=0 for j=0,1, and n=2 for j=2. Since the spatial dependence 
of $f_{n}(\vec{R})$ is known, we suppress the second index in the 
notation $\eta_{jn}$ and 
after transformations following that of Ref. \cite{Scharnberg80}, we arrive at a linear 
algebraic system for the amplitudes $\eta_{j}$:
\begin{eqnarray}
\lefteqn{(F_{00}+P-\ln 
\sqrt{h})\eta_{0}+\sqrt{2}(1+\epsilon)(F_{01}-P)\eta_{1}}\nonumber \\
&&\hspace{1cm}+ \frac{1-\epsilon^{2}}{\sqrt{2}}F_{02}\eta_{2}=0, 
\label{Sys1}\\
\lefteqn{\sqrt{2}(1+\epsilon^{2})(F_{01}-P)\eta_{0}
+(F_{00}-F_{01}+2P}\nonumber\\
&&\hspace{1cm}-\ln \sqrt{h}-\ln q)\eta_{1}+\frac{1-\epsilon}{2}
F_{02}\eta_{2}=0, \label{Sys2}\\
\lefteqn{\frac{1-\epsilon}{\sqrt{2}}F_{02}\eta_{0}+\frac{1-\epsilon^2}{2
}F_{02}\eta_{1}}\nonumber\\
&&\hspace{1cm}+(F_{00}+F_{01}-\ln \sqrt{h}-\ln q)\eta_{2}=0.
\label{Sys3}
\end{eqnarray}

The equations are written in the dimensionless units of Ref.  
\cite{Scharnberg80}: $h=\frac{2H}{H_{0}}$, $H_{0}=\frac{\Phi_{0}}{\pi 
\xi_{0}^2}$, $\xi_{0}=\frac{\hbar v_{F}}{2\pi T_{c0}}$, 
$\Phi_{0}=\frac{\pi \hbar c}{2}$, $t=\frac{T}{T_{c0}}$, $q=\frac{T_{c0}}{T_{c1}}$. 
For the calculation of angular averages, we assume a spherically 
symmetric Fermi surface so that $\vec{v}_{F}$ is parallel to 
$\vec{k}_{F}$.  Then the coefficients $F_{00}$, $F_{01}$, $F_{02}$ and 
$P$ are given by the following expressions:
\begin{eqnarray}
F_{00} & = & F_{00}\left(\frac{t}{\sqrt{h}}\right)=
\int^\infty_0\ln\left[\frac{\sqrt{h}}{t}\tanh\left(\frac{\rho 
t}{2\sqrt{h}}\right)\right] \nonumber\\
&&\left(\int^1_0(1-x^2)e^{-\frac{\rho^2}{4}(1-x^2)}dx\right)\frac{\rho 
d\rho}{2}\label{F0}\\
F_{01} & = & F_{01}\left(\frac{t}{\sqrt{h}}\right)=\int^\infty_0\ln
\left[\frac{\sqrt{h}}{t}\tanh\left(\frac{\rho t}{2\sqrt{h}}
\right)\right]\nonumber\\
&&\left(\int^1_0\frac{(1-x^2)(1-3x^2)}{2}e^{-\frac{\rho^2}{4}(1-x^2)}dx\right)
\frac{\rho d\rho}{2}\label{F1}\\
F_{02} & = & 
F_{02}\left(\frac{t}{\sqrt{h}}\right)=\frac{t}{4\sqrt{2h}}\int^\infty_0
\frac{\rho^2d\rho}{\sinh(\frac{\rho t}{\sqrt{h}})}\nonumber\\
&&\left(\int^1_0(1-x^2)e^{-\frac{\rho^2}{4}(1-x^2)}dx\right)\label{F2}\\
P & = & P(\frac{t}{\sqrt{h}},\lambda \sqrt{h})=-\frac{t}{\sqrt{h}}\int_{0}^{\infty}
\frac{1-\cos(\rho \lambda \sqrt{h})}{\sinh(\frac{\rho t}{\sqrt{h}})}\nonumber \\
&&\left(\int_{0}^{1}x^2e^{-\frac{\rho^{2}}{4}(1-x^{2})}dx\right) d\rho
\label{P}
\end{eqnarray}
with $\lambda = \frac{g\mu_{B}H}{2\pi T_{c0}}$.  $H_{c2}$(T) is found 
from the condition of compatibility of Eqns.  
(\ref{Sys1})-(\ref{Sys3}).  With the shorthand notations

\begin{center}$G = F_{00}-\ln \sqrt{h}-\frac{2}{3}Q$, where $Q=\ln 
q$,\end{center}

\noindent this condition has the form:
\begin{eqnarray}
\lefteqn{(G+\frac{2Q}{3})(G-\frac{Q}{3})^{2}-F_{01}^{2}(3G+2F_{01})+}\nonumber \\
&&3P[(G+F_{01})^{2}-\frac{Q^{2}}{9}]+\frac{9}{4}F_{02}^{2}(2F_{01}-3P-G)=0
\label{consistent}
\end{eqnarray}
which gives an implicit equation for $h$.


\section{Discussion}
Analytic solution of Eq.  (\ref{consistent}) is possible only in some 
limiting cases (cf.  Appendix).  In order to compare the predictions 
of Eq.  (\ref{consistent}) to the data, we have performed 
straightforward numerical calculations.  Fig.  
\ref{fig:differentes-RI} shows $H_{c2}(T)$ for three different 
irreducible representations together with the data on UBe$_{13}$. A 
pure $A_{1u}$ representation has a paramagnetic limitation 3 times 
higher than a pure (even) $A_{1g}$, leading to a good fit of the low 
field part of $H_{c2}(T)$ for a reasonnable value of the effective $g-factor$: 
$g=1.2$.  But the admixture of 
the $E_{u}$ representation is essential to reproduce the upturn of 
$H_{c2}(T)$ below $T_{c}/2$.

The surprise has been that this large influence of the admixture has 
been found for a parameter $T_{c1}/T_{c0}$ of only 0.12.  This comes 
from the fact that the admixture of the representations along the 
$H_{c2}$(T) curve is controled by the applied field, and it starts 
much above $T_{c1}$.  In any case, the corresponding admixture of 
$E_{u}$ components in the order parameter (defined in Eq.  
(\ref{fLandau})) remains less than 10\% down to T=0: see 
Fig.~\ref{fig:n1-n2}, where the coefficients $\eta_{1}$ and $\eta_{2}$ 
of definition (\ref{d123})-(\ref{fLandau}) have been reported normalized to 
$\sum\eta_{i}^{2}=1$.  So one is still far from the limit of 
complete suppression of the paramagnetic limitation (see Appendix).  
One can note on Fig.~\ref{fig:n1-n2} that most of the admixture of the 
$E_{u}$ representation comes from the $\vec{\Psi}_{1}$ function (Eq.  
(\ref{Eu1})), which compensates the $m_{z}=0$ component of the $A_{1u}$ 
representation with a Landau level $n=0$ instead of $n=2$ for the 
$\vec{\Psi}_{2}$ function (Eq.  (\ref{Eu2})).

A complete comparison of the best fits obtained in our model with the 
results under pressure of Ref. \cite{Glemot99} are 
presented on Fig. \ref{fig:Hc2}. Three parameters are used in the 
process of fitting: $v_{F}$, which is directly found from the slope 
$\left(-\frac{dH_{c2}}{dT}\right)_{T=T_{c}}$, the effective $g-factor$, which is 
determined by the curvature of the low-field part of the data, and the ratio 
$\frac{T_{c1}}{T_{c0}}$, which is controled by the high-field region.


The values of the fitting parameters for different pressures are 
presented on Fig. \ref{fig:param}. They suggest several comments.  
First of all, we note that the two-representations ($A_{1u}$, $E_{u}$) 
model can reproduce the unusual features of $H_{c2}(T)$ in UBe$_{13}$.  
Good quantitative agreement is obtained everywhere except in high 
fields.  At least two reasons can be invoked for the observed 
deviations: the effect of the omitted representations in equation 
(\ref{ScatAmp2}) is expected to become stronger in the low temperature 
range, and the strong coupling effects, known to exist in this 
compound (\cite{Ott87,WaltiPreprint}), have been neglected in our 
calculations and would also reinforce $H_{c2}(T)$ in this temperature 
range \cite{Glemot99}. So these deviations can be ascribed to some 
oversimplifications of our model, without fundamentally questionning 
its validity.

The fitting parameters have realistic values.  The pressure dependence 
of $v_{F}$ is in agreement with that of the Sommerfeld coefficient 
(deduced from Ref.  \cite{Phillips}).  This is not a surprise, as this 
feature is almost model independent and it has been already pointed 
out in Ref.  \cite{Glemot99}.  The effective $g-factor$ increases 
slowly from 1.2 to 1.5 and is not too far from the free electron 
value.  In the absence of any other probe of $g$, no additionnal 
cross-checking seems possible.

The more characteristic parameter of the model is the ratio 
$T_{c1}$/$T_{c0}$.  Pronounced effect of the subdominant interaction 
via admixture of the second representation is observed even 
when the temperature of the subdominant transition $T_{c1}$ is much 
smaller than the real transition temperature $T_{c0}$.  At ambient 
pressure $T_{c1}$/$T_{c0}$ is of the order of 0.1, whereas it grows 
linearly with pressure in the interval $0<P<20 kbar$, up to a value of 
0.7.  Extrapolation of this dependence in a region of higher pressures 
predicts a crossover of the two temperatures at $P \approx 30 kbar$.  
For $P>30 kbar$, $T_{c1}$ would be larger than $T_{c0}$ and the transition 
in zero magnetic field will take place in one of the three possible 
phases belonging to the $E_{u}$ representation \cite{Volovik85}.  An 
experimental observation of this crossover would be a good check of 
the proposed two-representation scheme.  The details of the crossover 
(if observed) could give support to the proposed choice of the two 
representations.


The observed qualitative changes (suppression of the upturn) of the 
curve $H_{c2}$(T) with increasing pressure are easily understood 
within the proposed scheme.  When $T_{c1}$ approaches $T_{c0}$, two 
representations can be considered as one three-dimensional 
representation, in which case the paramagnetic limitation can already be 
suppressed in low fields by a suitable combination of the basis 
functions, and no upturn will appear.  A possibility to fit better the experimental 
curve when $T_{c1}$ (and V$_{1}$) is growing is also natural, since in 
that case the role of the omitted representation is getting relatively 
smaller.

Eventually, let us note the difference between our model and the two 
transition proposal of ref.  \cite{Rauchschwalbe87}.  In the latter 
one, it is proposed that a second transition is present below $\approx 
0.6T_{c}$, and that the upturn comes from the apparition of a new 
phase with a weaker Pauli limitation.  The recent experimental effort 
reported in \cite{Kromer98} strives to detect such a transition 
already in zero field.  In our model, the second phase (the $E_{u}$ 
representation) does not appear in zero field-zero pressure.  It is 
only the applied field which introduces the mixture of the 
representations at finite temperatures (see Fig.  \ref{fig:n1-n2}).  A 
new phase transition is avoided precisely because the field allows 
such a mixture without additionnal symmetry breaking.  The present 
scenario, in addition to providing quantitative predictions, also has 
the advantage of giving a realistic account of the smoothness of all 
features observed below $T_{c0}$ in UBe$_{13}$.


\section{Conclusion}
The proposed scheme satisfactory explains the temperature dependence 
of $H_{c2}$(T) in UBe$_{13}$, but it is in conflict with the observed 
power-law temperature dependencies of the specific heat, London 
penetration depth and of the longitudinal relaxation time in 
NMR-experiments \cite{Ott87} which reveal the existence of gap nodes: 
with the $A_{1u}$ order parameter, the nodes in the superconducting 
gap can only be accidental.  The argument given above for the choice 
of the representations indeed leaves no other possibility than this 
$A_{1u}-E_{u}$ mixture among odd representations.  There remains a 
possibility of a mixture of even and odd representations 
\cite{Klemm85,Zhitomirskii92}.  The choice of possible pairs of 
competing representations in that case is much bigger.  We have not 
analyzed that possibility systematically though one would expect that 
in that case, because of a weaker coupling between the 
representations, the change in the curvature would be sharper.  A 
promising route could also be the inclusion of strong coupling effects, 
which have recently received microscopic experimental confirmation 
\cite{WaltiPreprint}

\acknowledgments
One of us (I.A.F.) is grateful to Universit\'e Joseph Fourier for a 
part-time position which made possible this work and to Prof.  J.  
Flouquet for his kind hospitality in the CEA Grenoble as well as for 
stimulating discussions.

\appendix
\section*{}
We consider now some limiting cases where the solution of Eq.  
(\ref{consistent}) with respect to $h$ can be found analytically.  This 
analysis is useful both for a clarification of the underlying physical 
picture as well as for a check of the numerical calculations.  Let us 
first consider the limit $T\rightarrow 0$.  The limiting values of the 
functions of the variable $t/\sqrt{h}$ $F_{00}$, $F_{01}$, $F_{02}$ 
for $t/\sqrt{h}\rightarrow 0$ are:
\begin{eqnarray}
F_{00}(0) &=& \int_{0}^{\infty}
\ln(\frac{\rho}{2})\frac{\rho d\rho}{2}
\int_{0}^{1}(1-x^2)e^{-\frac{\rho^{2}}{4}
(1-x^{2})}dx \nonumber \\
&=& \ln(\frac{e}{2\sqrt{\gamma}}) \label{AF}\\
F_{01}(0) &=& \int_{0}^{\infty}
\ln(\frac{\rho}{2})\frac{\rho d\rho}{4}\nonumber \\
&&\int_{0}^{1}(1-x^2)(1-3x^{2})
e^{-\frac{\rho^{2}}{4}(1-x^{2})}dx \nonumber \\
&=& -\frac{1}{6} \label{AF1}\\
F_{02}(0) &=& \frac{1}{4\sqrt{2}}\int_{0}^{\infty}\rho d\rho 
\int_{0}^{1}(1-x^2)e^{-\frac{\rho^{2}}{4}(1-x^{2})}dx \nonumber \\
&=& \frac{1}{3\sqrt{2}}
\label{AF2}
\end{eqnarray}

\noindent $P(\frac{t}{\sqrt{h}},\lambda \sqrt{h})$ in a limit 
$t/\sqrt{h}\rightarrow 0$ remains a function of $\lambda \sqrt{h}$:
\begin{eqnarray}
P&=&P(0,\lambda \sqrt{h}) \nonumber\\
&=&-\int_{0}^{\infty}
\frac{d\rho}{\rho}[1-\cos(\rho \lambda \sqrt{h})]
\int_{0}^{1}x^2e^{-\frac{\rho^{2}}{4}(1-x^{2})}dx
\label{AP}
\end{eqnarray}
For a further simplification we consider the limit $\lambda 
\sqrt{h}\rightarrow \infty$ as being relevant to the actual situation 
in UBe$_{13}$.  To evaluate the asymptotics of the integral in Eq.  
(\ref{AP}) in that limit, let us split the interval of integration in 
two: $(0,\rho_{0})$ and $(\rho_{0},\infty)$ where $\rho_{0}$ is chosen 
to meet the following condition:
\begin{equation}
\frac{1}{\lambda \sqrt{h}} \ll \rho_{0} \ll 1
\label{Arho0}
\end{equation}
In the first interval, one can then assume 
$e^{-\frac{\rho^{2}}{4}(1-x^{2})} \approx 1$, and the contribution 
$I_{1}$ of this interval is evaluated straightforwardly:
\begin{equation}
I_{1}=\frac{1}{3}\int_{0}^{\rho_{0}}
\frac{d\rho}{\rho}[1-\cos(\rho \lambda \sqrt{h})]
=\frac{1}{3}\ln(\gamma \lambda \sqrt{h} \rho_{0}),
\label{AI1}
\end{equation}
where $\ln\gamma = C$ is Euler's constant. When integrating over the 
second interval $(\rho_{0},\infty)$, one can drop the oscillating term 
$cos(\rho \lambda \sqrt{h})$, yielding for the contribution 
$I_{2}$ of this interval:

\begin{center}$I_{2}=\int_{0}^{1}
\frac{x^2dx}{2}\int_{u_{0}}^{\infty}\frac{du}{u}e^{-u}$, 
where $u_{0}=(1-x^{2})\frac{\rho_{0}^{2}}{4}$\end{center}

\noindent Integrating over u by parts and taking into account that
\begin{equation}
\int_{0}^{\infty}\ln u e^{-u} du = -C = -\ln \gamma \nonumber
\end{equation}
we arrive at the following contribution of $I_{2}$:
\begin{equation}
I_{2}=-\frac{1}{3}\ln(\rho_{0}\sqrt{\gamma}) + \frac{4}{9}
\label{AI2}
\end{equation}
The sum of (\ref{AI1}) and (\ref{AI2}) gives the principal order terms in the
asymptotics of P for $\lambda \sqrt{h}\rightarrow \infty$:
\begin{equation}
P=\frac{1}{3}\ln(\lambda \sqrt{\gamma h}) + \frac{4}{9}
\label{APfinal}
\end{equation}
Given the limiting values Eqs.~(\ref{AF}) - (\ref{AF2}) and the asymptotic 
(\ref{APfinal}), one can find a limiting value of $H_{c2}$ at T=0 and 
$\lambda \sqrt{h}\rightarrow \infty$.  According to (\ref{APfinal}) $P 
\rightarrow \infty$ when $\lambda \sqrt{h}\rightarrow \infty$.  
Collecting in Eq. (\ref{consistent}) the terms proportional to P and setting them 
to zero we arrive at the following equation:
\begin{equation}
(G+F_{01})^{2} = \frac{Q^{2}}{9} + \frac{9}{4}F_{02}^{2}
\label{Aconsistent}
\end{equation}
Using here the limiting values Eqs.~(\ref{AF}) - (\ref{AF2}) we solve this 
equation with respect to h. The largest of the two roots gives $H_{c2}$:
\begin{equation}
h = \frac{1}{4\gamma q^{4/3}}\exp\left[\frac{5}{3} + 2\sqrt{\frac{Q^{2}}{9} + 
\frac{1}{8}}\right]
\label{AHc2}
\end{equation}
For this value of h one obtains from Eqns. (\ref{Sys1}) - 
(\ref{Sys3}) the limiting values of the ratios of the coefficients in
the definition (\ref{fLandau}) :
\begin{equation}
\frac{\eta_{1}}{\eta_{0}} = \frac{1+\epsilon^{2}}{\sqrt{2}}
\label{Aratio1}
\end{equation}
This ratio corresponds to a complete elimination of the z-projection 
of the order parameter (cf. Eq. (\ref{PsiInfini})).

Another ratio is:
\begin{equation}
\frac{\eta_{1}}{\eta_{2}} = \frac{2\sqrt{2}}{3}(1-\epsilon)
\left[\frac{Q}{3} + \sqrt{\frac{Q^{2}}{9} + \frac{1}{8}}\right] 
\label{Aratio2}
\end{equation}

The order parameter obtained with these coefficients has the form:
\begin{equation}
\vec{d}(\vec{k}) = \sqrt{3}\hat{y}k_{y}
\label{dlimite}
\end{equation}

In the opposite limit when $T\rightarrow T_{c}$ and $H_{c2}\rightarrow 
0$ we can expand the functions entering Eq. (\ref{consistent}) in 
powers of h and $\tau=1-t$.  These functions are defined as double 
integrals.  For integration over $\rho$ the convergence is provided by 
the exponential factor $e^{-\frac{\rho t}{\sqrt{h}}}$ in a region 
$\rho \sim \sqrt{h}/t \ll 1$.  It means that all other functions can be 
expanded in powers of $\rho^2$ which, after integration over $\rho$, gives 
an
expansion over $h/t^2$.  We keep only terms which are necessary for finding 
$\tau (h)$ with an accuracy up to h$^{2}$:\\

$F_{00} - \ln \sqrt{h}= -\ln t -  a\frac{h}{t^2} + 
b\frac{h^2}{t^4}$, $F_{01} = - f_{1}\frac{h}{t^2},$\\

$F_{02} = f_{2}\frac{h}{t^2}$, $P= -a \lambda^2 \frac{h^2}{t^2}$,\\

\noindent with $a=\frac{7}{12}\zeta (3)$, $b=\frac{31}{40}\zeta (5)$, 
$f_{1}=\frac{7}{60}\zeta (3)$, $f_{2}=\frac{7}{15\sqrt{2}}\zeta (3)$ 
where $\zeta (z)$ is Riemann's zeta function. Substitution of this 
expansion in Eq. (\ref{consistent}) gives
\begin{equation}
\tau (h) = ah+(\frac{3}{2}a^2-b+a\lambda^2-
\frac{1}{Q}[2f_{1}^2+\frac{3}{2}f_{2}^2])h^2
\label{Atau}
\end{equation}

The term linear on $h$ comes from $F_{00}$ and is determined entirely by 
the $A_{1u}$ representation, but the $h^{2}$ term is influenced by 
the $E_{u}$ representation via the functions $F_{01}$ and $F_{02}$ (or 
the coefficients $f_{1}$ and $f_{2}$). This influence is getting stronger 
when $Q$ decreases, i.e. when the two transition temperatures are coming 
closer. The term proportional to $1/Q$ is definitely negative: it 
means that for sufficiently small $Q$, the dependence of $H_{c2}(T)$ 
will have positive curvature starting from $T_{c}$.


\begin{figure}[]
\caption{Upper critical field of UBe$_{13}$ at zero pressure (from 
\protect\cite{Thomas96}), together with three 
calculations of $H_{c2}(T)$ for three different 
irreducible representations: $A_{1g}$, $A_{1u}$, and a mixture 
$A_{1u}-E_{u}$. The same values of the effective $g-factor$ (adjusted for 
the odd parity representations) and the Fermi velocity have been used 
for the three fits.}
\label{fig:differentes-RI}
\end{figure}

\begin{figure}[]
\caption{Temperature dependence of the normalized components 
$\eta_{1}$ and $\eta_{2}$ of the respective basis functions 
$\vec{\Psi}_{1}$ and $\vec{\Psi}_{2}$ of the $E_{u}$ representation, 
along the $H_{c2}(T)$ fit of Fig \protect{\ref{fig:differentes-RI}}.  
Note that they both start to grow much above $T_{c1}=0.12 T_{c0}$.}
\label{fig:n1-n2}
\end{figure}

\begin{figure}[]
\caption{Upper critical field of UBe$_{13}$ under pressure 
\protect\cite{Thomas96}, and best fit of our $A_{1u}-E_{u}$ model (full 
lines) for each pressure: the main features are well reproduced, and 
deviations appear only at low temperatures (see discussion in the 
main text).}
\label{fig:Hc2}
\end{figure}

\begin{figure}[]
\caption{Pressure (p) dependence of the three parameters of the fits 
of Fig.  \protect{\ref{fig:Hc2}}.  The comparison of the 
p-dependence of the Fermi velocity $v_{F}$ with the Sommerfeld 
coefficient $\gamma (p)$ of ref.  \protect\cite{Phillips} is 
gratifying.  Extrapolation of the linear increase of 
$\frac{T_{c1}}{T_{c0}}$ predicts that the $E_{u}$ representation will 
appear first in zero field above $p\approx 30 kbar$.}
\label{fig:param}
\end{figure}

%
%
%
%

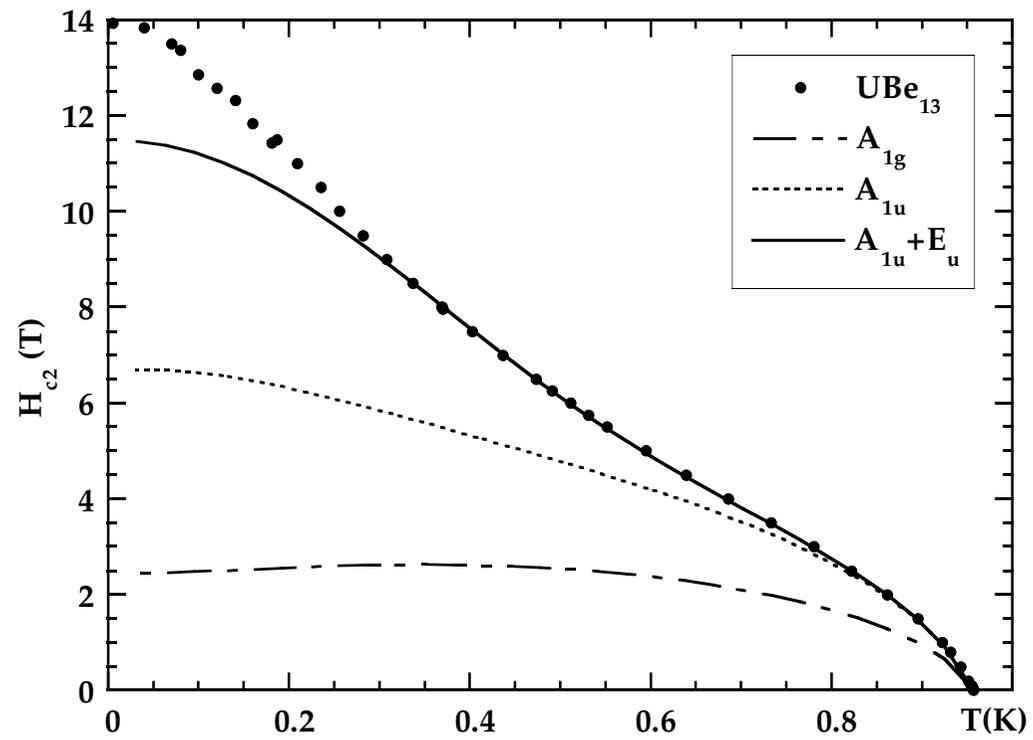
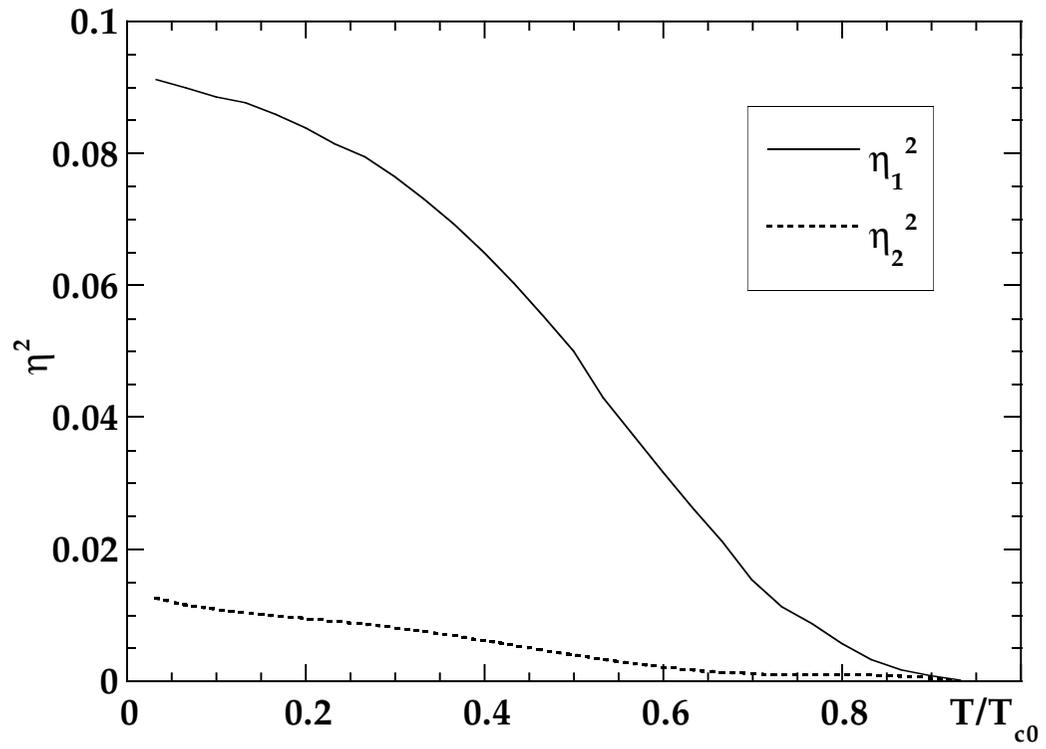
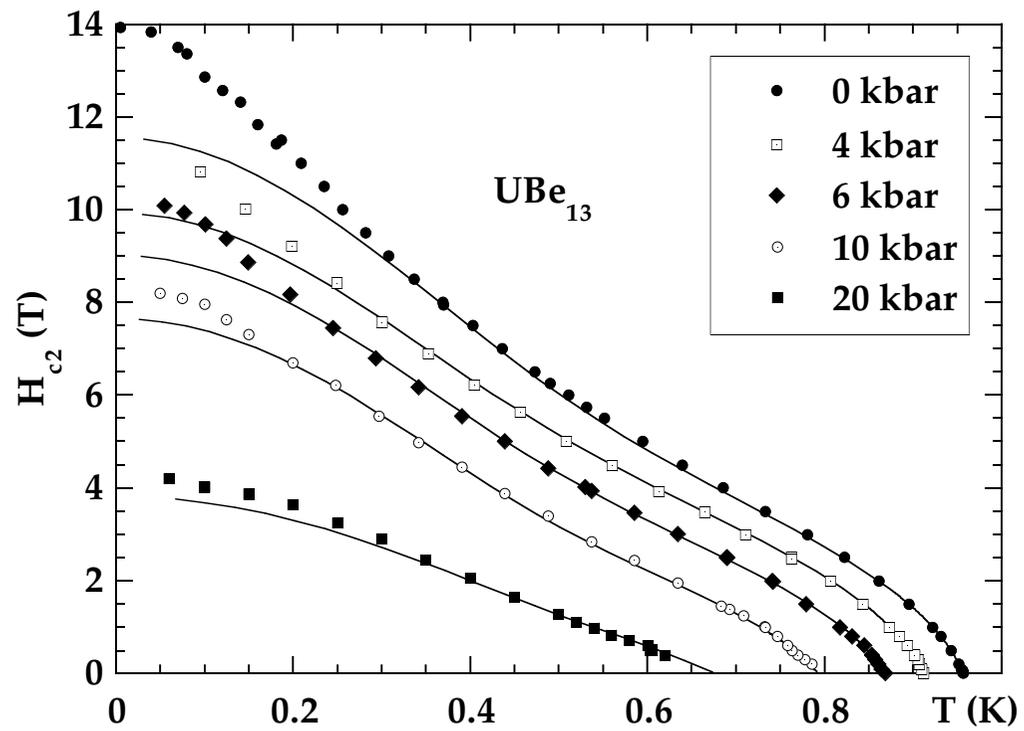
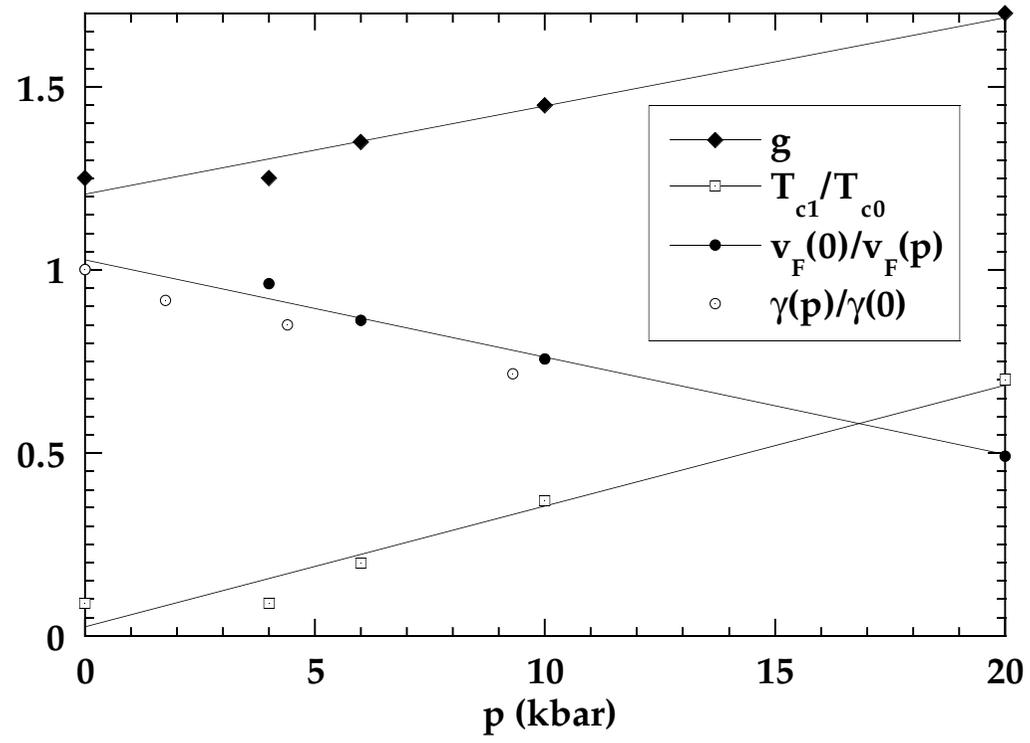
\end{document}